\documentclass[letterpaper, 10 pt, conference]{ieeeconf}  
\IEEEoverridecommandlockouts                              

\usepackage{cite}
\usepackage{amsmath,amssymb,amsfonts,amsbsy}
\usepackage{algorithmic}
\usepackage{graphicx}
\usepackage{textcomp}
\usepackage{xcolor}
\usepackage{array}
\usepackage{float}
\usepackage{multirow}
\usepackage{times}
\usepackage{comment}
\usepackage{url,theorem}
\usepackage{comment}
\usepackage{svg}

\newtheorem{Proposition}{\textbf{Proposition}}
\newtheorem{definition}{\textbf{Definition}}
\newtheorem{remark}{\textbf{Remark}}
\newtheorem{Algorithm}{\textbf{Algorithm}}

\newcommand{\qed}{\hfill$\Box$}
\newcommand{\Fn}{\mathrm{F}}
\newcommand{\Hn}{\mathrm{H}}
\newcommand{\Tn}{\mathrm{T}}

\allowdisplaybreaks

\title{\LARGE \bf
A frequency-domain approach for estimating continuous-time diffusively coupled linear networks
}

\author{Desen Liang, E.M.M. (Lizan) Kivits, Maarten Schoukens and Paul M.J. Van den Hof
\thanks{Funded by the European Union. Views and opinions expressed are however those of the author(s) only and do not necessarily reflect those of the European Union or the European Research Council. Neither the European Union nor the granting authority can be held responsible for them.}
\thanks{The authors are with the Control Systems Group, Department of Electrical Engineering, Eindhoven University of Technology, The Netherlands.
{\tt\small d.liang.cn@outlook.com, \{e.m.m.kivits, m.schoukens, p.m.j.vandenhof\}@tue.nl}.}
}

\begin{document}
\maketitle
\thispagestyle{empty}
\pagestyle{empty}

\begin{abstract}
This paper addresses the problem of consistently estimating a continuous-time (CT) diffusively coupled network (DCN) to identify physical components in a physical network. We develop a three-step frequency-domain identification method for linear CT DCNs that allows to accurately recover all the physical component values of the network while exploiting the particular symmetric structure in a DCN model. This method uses the estimated noise covariance as a non-parametric noise model to minimize variance of the parameter estimates, obviating the need to select a parametric noise model. Moreover, this method is extended to subnetworks identification, which enables identifying the local dynamics in DCNs on the basis of partial measurements. The method is illustrated with an application from In-Circuit Testing of printed circuit boards. Experimental results highlight the method's ability to consistently estimate component values in a complex network with only a single excitation.
\end{abstract}

\setcounter{page}{1}

\section{INTRODUCTION}
Physical dynamic networks consist of interconnections of physical components, which can be described by diffusive couplings. They can model various processes in numerous fields, such as electrical circuits, mechanical machines, and chemical processes\cite{Dorfler}. System identification uses statistical methods to construct mathematical models of dynamical systems from measured data \cite{soderstrom1989system} and is widely used in fault detection and diagnosis (FDD) \cite{milijkfaultdetection}. As physical systems such as printed circuit board assemblies (PCBAs) grow in complexity, there is an increasing interest in physical network identification to effectively identify the dynamics of interconnected systems. One application of this technique is In-Circuit Testing, which diagnoses faults in PCBAs by utilizing data from test probes. As PCBAs can be modeled as physical dynamic networks, the idea is to estimate each electronic component value with the network structure, compare it with the expected value, and analyze the difference in values to verify its correctness.

Several data-driven methods are available to identify physical components in networks. For example, physical systems can be identified by estimating structured state-space models \cite{Mercere2014}\cite{YU201854}; physical systems can also be treated as directed dynamic networks with specific structural constraints in \cite{lizan2019}, where prediction error methods are used for estimation.
As a more direct approach, a DCN model has been developed in \cite{lizan} where the physical system is represented as a non-directed network, and a discrete-time (DT) multistep algorithm has been developed to identify structured polynomial models of DCNs incorporating the network structure. The structured polynomials directly represent the physical component values of the network. This algorithm has later been applied to subnetworks \cite{lizan2022}. However, since physical component values are most often represented in the CT domain, there is a need for developing an estimation algorithm for estimating a continuous-time model in the same DCN setting.  This will be the topic of the current paper.

Numerous estimation methods involving instrumental variables have been used to identify CT models \cite{garnier_new_2021}\cite{Gonzalez2023}. Indirect and direct CT identification approaches for dynamic networks that are modeled as interconnected CT transfer functions have been developed in \cite{dank2014}. Frequency domain-based techniques have been applied to identify CT and DT models directly within the same algorithm. Alternatively, CT systems with multiple inputs and multiple outputs (MIMO) can be identified using a two-step frequency-domain approach by first estimating the non-parametric frequency response functions (FRF) and a frequency-dependent noise covariance, and then using these to identify parametric transfer function models \cite{pintelon2010estimation1, pintelon2010estimation-2, pintelon2012system}. Utilizing these two-step methods, a local module identification method in dynamic networks has been proposed in \cite{ramaswamy2022frequency}. As the physical component values are intertwined within the transfer function's coefficients, it is preferable and straightforward to retrieve these values from the well-structured polynomial matrix of the DCN model. Therefore, in this paper, we aim to develop a method that directly identifies CT DCNs while preserving the physical network structure, enabling accurate estimation of the network's physical components.

The research question we aim to answer in this paper is: \textit{How can we consistently identify the physical component values in a DCN model in CT?} This paper explores two scenarios: first, is to identify all the dynamics in a CT DCN (full network identification), and second, is to identify the local dynamics in a CT DCN (subnetwork identification). This paper addresses the research question by directly identifying a CT DCN model in the frequency domain. An identification procedure is developed by extending the frequency-domain approach in \cite{pintelon2012system} to DCNs for full network identification, and is further extended to identify local dynamics with partial measurements for subnetwork identification. The paper demonstrates consistent estimation results for both full network and subnetwork identifications.

After presenting the DCN and its frequency-domain model in Section~\ref{Preliminary}, we present a three-step frequency-domain identification algorithm in Section~\ref{Sec:fdomain identificaiton algorithm}. This algorithm is extended to identify the subnetworks with partial measurements in Section~\ref{sec:local id}. The algorithms will be applied to In-Circuit testing experiments, of which the results are shown and discussed in Section~\ref{sec:FDD}. The conclusion is provided in Section~\ref{sec:conclusion and future}.

Consider the following notation throughout the paper. $A(p)$ is a polynomial matrix with $A(p)=\sum_{\ell=0}^{{{n}_{a}}}{{{A}_{\ell}}{{p}^{\ell}}}$, it consists of $n_a+1$ matrices $A_\ell$ and $(j,k)$-th polynomial elements $a_{jk}(p) = \sum_{\ell=0}^{{{n}_{a}}}{{{a}_{jk,\ell}}{{p}^{\ell}}}$. ${a}_{jk,\ell}$ is the $(j,k)$-th element of the matrix $A_\ell$.

\section{DIFFUSIVELY COUPLED NETWORKS}\label{Preliminary}

\subsection{Continuous-time model}
A linear diffusive coupling describes an interconnection among node signals in which the coupling strength is proportional to the signal difference between the nodes. This coupling yields symmetric interactions for the linear components. We can describe the full dynamics and topology of linear DCNs as follows according to \cite{lizan}.
\begin{definition}(DCN)\label{def:diffusively network}
  A DCN consisting of $L$ node signals, collected in $w(t)$, and $K$ excitation signals $r(t)$, is described as
  \begin{equation}
    A(p)w(t)=B(p)r(t)+F(p)e(t),
    \label{eq:def-idmodel}
  \end{equation}
  where $p$ is the differential operator, i.e., $p^\ell w_j(t)=w_j^{(\ell)}(t)$, $w_j^{(\ell)}(t)$ is the $\ell$-th order derivative of node signals $w_j(t)$,
  \begin{itemize}
    \item $A(p)=\sum_{\ell=0}^{{{n}_{a}}}{{{A}_{\ell}}{{p}^{\ell}}\in {{\mathbb{R}}^{L\times L}}[p],\text{with }{{a}_{jk}}(p)={{a}_{kj}}(p)}$,
    ${\forall k,j\text{, and }{{A}^{-1}}(p)\text{ is stable}\text{.}}$
    \item $B(p)\in {{\mathbb{R}}^{L\times K}}[p]$.
    \item $F(p)\in {{\mathbb{R}}^{L\times L}}(p)$ is monic, stable, rational, and the inverse matrix is also stable.\qed
  \end{itemize}
\end{definition}

The DCN is assumed to be connected, which means that there is a path between each pair of nodes. The polynomial matrix $A(p)$ is symmetric and non-monic, capturing the symmetric diffusive couplings in the system. The polynomial matrix $B(p)$ characterizes the dynamics through which the excitation signals $r(t)$ enter the network and is chosen in this paper as binary and known. To model the unknown disturbance signals, $F(p)$ is a rational matrix that captures the noise dynamics, and $e(t)$ represents a wide-sense stationary white noise process that is independent and identically distributed (i.i.d.). The input/output signals follow the band-limited measurement assumption.
\begin{remark}
  Note that pre-multiplying \eqref{eq:def-idmodel} with $A^{-1}(p)$ yields a transfer function form. If $F(p)$ is a polynomial or identity matrix, the model \eqref{eq:def-idmodel} yields a non-monic ARMAX-like or non-monic ARX-like model structure, respectively. In this paper, $F(p)$ is chosen as a polynomial matrix, resulting in a nonmonic ARMAX-like model structure.\qed
\end{remark}

We can reformulate \eqref{eq:def-idmodel} by decomposing $A(p)$ into $X(p)+Y(p)$ as
\begin{equation}
  X(p)w(t)+Y(p)w(t)=B(p)r(t)+F(p)e(t),
  \label{matrix form for A}
\end{equation}
where $X(p)$ is a diagonal polynomial matrix which represents the grounded dynamics of the DCN; $Y(p)$ is a Laplacian\footnote{A Laplacian matrix is a symmetric matrix with nonpositive off-diagonal elements and with nonnegative diagonal elements that are equal to the negative sum of all other elements in the same row (or column) \cite{Mesbahi2010}.} polynomial matrix of which the off-diagonal elements represents the interconnected dynamics of the DCN. The symmetric polynomial matrix $A(p)=X(p)+Y(p)$ is used to represent the dynamics of the DCN through which $X(p)$ and $Y(p)$ can be uniquely recovered.

\begin{figure}[tbp]
    \vspace{-10pt}
    \centering
    \includegraphics[width=\columnwidth]{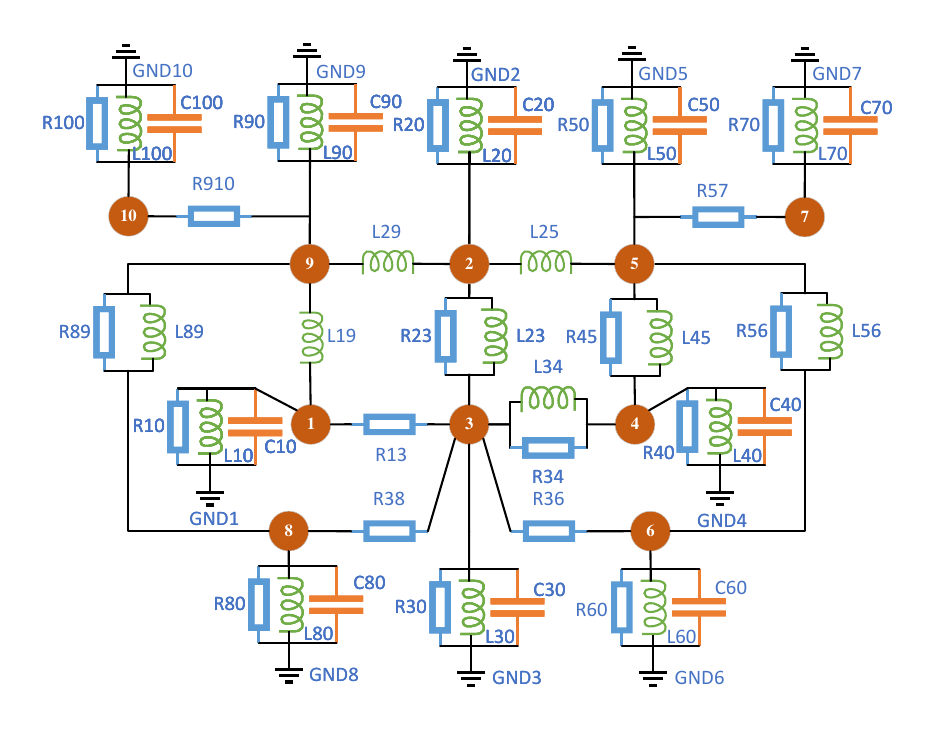}
    \vspace{-30pt}
    \caption{A 10-node RLC circuit with inductors ($L_{jk}$), resistors ($R_{jk}$), capacitors ($C_{jk}$), and ground nodes($GND_{j}$).}
    \label{fig:RLC circuit 10nodes}
    \vspace{-20pt}
\end{figure}

For example, in the 10-node RLC circuit of Fig.~\ref{fig:RLC circuit 10nodes}, $X_{33}(p)$ contains the dynamics of $C_{30}$, $R_{30}$, and $L_{30}$; while $Y_{23}(p) = Y_{32}(p)$ contains the dynamics of $R_{23}$, and $L_{23}$.

\subsection{Frequency-domain model}\label{subsec:Frequency domain model}
Consider the DCN system described in \eqref{eq:def-idmodel} where CT signals $w(t)$, $r(t)$, and $e(t)$ are measured and sampled in time intervals $t_n = n_s T_s$ ($n_s = 0,...,N-1$) with a sampling time $T_s$.
The Discrete Fourier Transform (DFT) is used to transform the time-domain model to the frequency-domain. The DFT samples $S(k)$ for the signal $s(t_n)$ are defined as
\begin{equation}
  S(k)=\frac{1}{\sqrt{N}} \sum_{t_n=0}^{N-1} s(t_n) e^{-j 2 \pi k t_n / N}.
  \label{eq:DFT}
\end{equation}
Applying this DFT to the signals $w(t_n)$, $r(t_n)$, and $e(t_n)$ results in samples $W(k)$, $R(k)$, and $E(k)$, respectively. Accordingly, the DCN frequency-domain model is obtained.

\begin{definition}(Frequency-domain model)\label{def:frequency network model}
  The description of the DCN model in the frequency-domain is defined as
  \begin{equation}
    A(\Omega_k)W(k)=B(\Omega_k)R(k)+F(\Omega_k)E(k)+C(\Omega_k)
    \label{eq:frequency domain_k}.
  \end{equation}
  Pre-multiplying with $A^{-1}(\Omega_k)$ on both side gives
  \begin{multline}
    W(k)=\underbrace{{{A}^{-1}(\Omega_k)}B(\Omega_k)}_{G(\Omega_k)}R(k)+\underbrace{{{A}^{-1}(\Omega_k)}F(\Omega_k)}_{H(\Omega_k)}E(k)\\
    +\underbrace{{{A}^{-1}(\Omega_k)}C(\Omega_k)}_{T(\Omega_k)},
    \label{eq:frequency domain_k the classical one}
  \end{multline}
  where $\Omega_k$ is the frequency variable for sample $k\text{ }(k=1,...,N)$ and defined as $\Omega_k=j\omega_k$ in CT and $\Omega_k=e^{j\omega_kT_s}$ in DT, $\omega_k=\frac{2\pi kf_s}{N}$ and $f_s = \frac{1}{T_s}$ is the sampling frequency; $G(\Omega_k) \in \mathbb{C}^{L\times K}$ is the input/output dynamics and $H(\Omega_k)\in \mathbb{C}^{L\times L}$ is the noise model; $T({{\Omega }_{k}})\in \mathbb{C}^{L\times 1}$ is the transient term including the system and noise transient in time domain, which cause leakage errors in the frequency domain \cite{pintelon2010estimation1}. \qed
\end{definition}

Moreover, for using the non-parametric noise model later in the paper, the $\Omega_k$-dependent noise covariance is defined as \begin{equation}
      C_V(k) = Cov(V(k)), \text{with } V(k) = H(\Omega_k)E(k).
  \end{equation}


\section{FREQUENCY-DOMAIN IDENTIFICATION}\label{Sec:fdomain identificaiton algorithm}
In this section a three-step frequency-domain identification approach is presented to consistently estimate the physical components in a DCN. The FRF and frequency-dependent noise covariance are estimated in the first step using a non-parametric methodology. In the second step a parametric DCN model is estimated on the basis of the FRF and a non-parametric noise model, through linear regression steps in a Sanathan-Koerner-type of iteration. In the third step this parametric model is refined through a (nonconvex) maximum likelihood estimation (MLE). This approach offers two main advantages: $(1)$ avoiding the noise model parameterization and the need to determine its structure and order; $(2)$ with the transient term removed in the non-parametric model, the FRF estimation serves to extract ``transient-free" input/output data, facilitating the parametric identification.

\subsection{Step 1: non-parametric identification}\label{sec:non-parameteric}

The first step is to estimate the FRF $G(\Omega_k)$ and the frequency-dependent noise covariance $C_V(k)$. We will use the Local Polynomial Method (LPM), which is an advanced approach that accurately estimates the non-parametric FRF and noise covariance matrix without being restricted to use periodic excitation \cite{pintelon2012system}. In this paper, LPM is implemented using the `FrequencydomainToolbox' in \cite{pintelon2012system}.

Following the classical LPM method\cite{pintelon2012system}, we estimate polynomials over a short frequency range to obtain a local smooth least-squares approximation of the frequency functions. The FRF $G(\Omega_k)$ and the transient term $T(\Omega_k)$ of the network model \eqref{eq:frequency domain_k the classical one} are approximated around the selected central frequency $k$ in each frequency band $k+r$ with $r = -n,-n+1,...0,...,n-1,n$,
\begin{align}
  W(k+r)=&\left[G\left(\Omega_k\right)+\sum_{s=1}^\tau g_s(k) r^s\right]R(k+r)\nonumber\\
  &\hspace{1em}+T\left(\Omega_k\right)+\sum_{s=1}^\tau t_s(k) r^s+V(k+r),\\
  =&\Theta Z(k+r)+V(k+r),
  \label{eq:frequency domain_k+r with Theta}
\end{align}
where $\tau$ is the order of the polynomial; $L\times (\tau+1)(K+1)$ matrix $\Theta$ collects all the polynomial coefficients as $\Theta =[\begin{array}{cccccccc}
G\left(\Omega_k\right)&g_1(k) &... &g_\tau(k) &T\left(\Omega_k\right)&t_s(k)&... &t_\tau(k)
\end{array}]$; and $(\tau+1)(K+1)\times1$ vector $Z$ collects the input data.
Collecting \eqref{eq:frequency domain_k+r with Theta} in the frequency band $k+r$ ($2n+1$ samples) at central frequency $k$ and stacking them in a matrix gives
\begin{equation}
  W_n = \Theta Z_n + V_n,
  \label{eq:matrix version frequency}
\end{equation}
where, $W_n$, $Z_n$ and $V_n$ are $L\times(2n+1)$, $(\tau+1)(K+1)\times(2n+1)$, and $L\times(2n+1)$ matrices, respectively. The parameter matrix estimate $\hat{\Theta}$ at central frequency $k$ is obtained by minimizing a least squares cost function locally
\begin{equation}
  \hat{\Theta}= \arg\min _{\Theta}\left||W_n-\Theta Z_n|\right|^2_\Fn,
  \label{eq:LPM-LS}
\end{equation}
where, for any matrix $M$, $\left||M|\right|_\Fn$ is the Frobenius norm of matrix $M$.   The optimization \eqref{eq:LPM-LS} is solved by using a numerically stable method (the singular value decomposition). The estimated $L\times K$ matrix $\hat{G}(\Omega_k)$ is extracted from $\hat{\Theta}$
\begin{equation}
  \hat{G}\left(\Omega_k\right)=\hat{\Theta}_{\left[:, 1: K\right]}=W_nZ_n^\Hn(Z_nZ_n^\Hn)^{-1}_{\left[:, 1: K\right]},
  \label{eq:estimateFRF}
\end{equation}
where, for any complex matrix $M$, $M^\Hn$ is the Hermitian conjugate transpose of $M$; $\hat{\Theta}_{\left[:, 1: K\right]}$ represents the first $K$ columns extracted from the matrix $\hat{\Theta}$. Finally, substituting the estimation of the polynomial coefficients into \eqref{eq:matrix version frequency}, we can get the non-parametric noise estimation,
\begin{equation}
  \hat{V}_n=W_n-\hat{\Theta}Z_n = W_n[I_{2n+1}-Z_n^\Hn(Z_nZ_n^\Hn)^{-1}Z_n].
\end{equation}
 With this, the estimated noise covariance $\hat{C}_V(k)$ is obtained from the residual of \eqref{eq:LPM-LS} in line with \cite{pintelon2010estimation1},
\begin{equation}
  \hat{C}_V(k)=\frac{\hat{V}_n\hat{V}_n^\Hn}{2n+1-(\tau+1)(K+1)}.
  \label{noise covariance}
\end{equation}
By moving the frequency band $k+r$ for all frequencies, we can estimate the FRF and noise covariance for the whole range of frequencies.
\begin{remark}
  The order $\tau$ of the polynomials is chosen as 3 to compromise leakage and interpolation error. The quality of the noise model depends only on the frequency bandwidth $2n+1$ and the order of the local polynomial approximation $\tau$. Using this non-parametric noise model avoids performing a model order selection process as is required for classical PEM-based parametric identification.\qed
\end{remark}

\subsection{Step 2: structured polynomial matrix identification}\label{sec:parametric id}
In this step, the model set is defined as a collection of parameterized models that are considered candidates to represent the behavior of the system; the identification conditions are presented for consistent identification; and a constrained convex optimization method is considered to identify a parametric network model.

For estimating a parametric model based on $\hat{G}$ we select a model set,
\begin{equation}
  \mathcal{M} = \{(A(p,\theta),B(p,\theta),C(p,\theta)), \theta \in \Psi \subset\mathbb{R}^d\},
\end{equation}
with $d\in \mathbb{N}$, where $\theta$ includes all unknown parameters of the DCN model $A(p), B(p)$ and $C(p)$. Besides, a data generating system $\mathcal{S}$ is defined as,
\begin{equation}
    \mathcal{S}=(A^0(p,\theta^0),B^0(p,\theta^0),C^0(p,\theta^0),F^0(p,\theta^0)).
\end{equation}
The condition that data generating system belongs to the considered model set is needed for a consistent estimate. Since the noise model is not parameterized, the noise model is independent of the plant model. We only need $P_{0} \in \mathcal{M}$ for a consistent estimate in this paper, with $P_{0} = \left(A^0(p,\theta^0),B^0(p,\theta^0),C^0(p,\theta^0)\right)$.

\begin{Proposition}
  The DCN is identifiable if the following conditions are satisfied according to \cite{lizan} and \cite{lizanidentifibility}:
  \begin{enumerate}
    \item The polynomials $A(p)$ and $B(p)$ are left-coprime.
    \item There exists a permutation matrix $P$ that leads to $\left[A_0\text{ }A_1\text{ }...\text{ }A_{n_a}\text{ }B_0\text{ }B_1\text{ }...\text{ }B_{n_b}\right] P =\left[D \text{ }U\right]$ with $D$ square, diagonal, and full rank.
    \item There exists at least one excitation signal, i.e., $K\geq1$.
    \item There exists at least one constraint on the parameters of $A(p,\theta)$ and $B(p,\theta)$ that ensures $\Gamma \theta = \gamma \neq 0$, where $\Gamma$ is a full-row rank matrix.\label{constraint_con} \qed
  \end{enumerate}
  \label{condition:para}
\end{Proposition}

\begin{remark}
  The first condition ensures that there are no common factors between $A(p,\theta)$ and $B(p,\theta)$. The non-monicity of the polynomial matrix $A(p,\theta)$ is addressed in the second and fourth conditions, guaranteeing the uniqueness of this model representation.\qed
\end{remark}

As the transfer function $G(\Omega_k,\theta)$ is defined as the ratio of two polynomial matrices in \eqref{eq:frequency domain_k the classical one}, a commonly used method to identify this parametric $G(\Omega_k,\theta)$ would be a Gaussian-Newton (GN) based method in which the following nonlinear output error criterion is minimized,
\vspace{-5pt}
\begin{equation}
    \hat{\theta }=\arg\underset{\theta }{\mathop{\min }}\,\frac{1}{N}\sum\limits_{k=1}^{N}||W(k)-G(\Omega_k,\theta)R(k)-T(\Omega_k,\theta)||^2_\Fn.
    \label{eq:nonliear_criterion}
\end{equation}
However, there is no guarantee of reaching the global minimum in such non-convex optimization, possibly leading to inaccurate physical component estimates in the network. To address this problem, we follow the idea of the SK-iteration algorithm \cite{sanathanan1963transfer}, determined by
\begin{align}
 \hat\theta^i =& \arg\min_{\theta}\frac{1}{N}\sum\limits_{k=1}^{N}
 \|A(\Omega_k,\hat\theta^{i-1})^{-1} \cdot \\
 &  [A(\Omega_k,\theta)W(k)-B(\Omega_k,\theta)R(k)
    -C(\Omega_k,\theta)] \|^2_\Fn .
    \label{eq:SK_iter_ori}
\end{align}
When the algorithm converges, its stationary point will be close to the global minimum. An additional frequency weighting can be applied in the criterion \eqref{eq:SK_iter_ori} through the non-parametric noise model $\hat{C}_V(k)^{\frac{1}{2}}$ in each iteration as another frequency weighting to reduce variance. However, since $\hat{C}_V(k)$ is obtained from the residual in \eqref{eq:LPM-LS}, $\hat{C}_V(k)$ and $W(k)$ are not independently distributed, leading to an inconsistent estimation.

As an alternative, we employ the asymptotically independent distributed sample covariance and sample mean, which are derived from the LPM estimation of the node signals $W(k)$ and its associated covariance according to \cite{pintelon2012system}. Utilizing the non-parametric estimate matrix $\hat{\Theta}$ in \eqref{eq:LPM-LS}, the sample mean $\hat{W}(k)$ of $W(k)$ is calculated as
\begin{equation}
  \hat{W}(k)=\hat{G}\left(\Omega_k\right) R(k)+\hat{T}\left(\Omega_k\right),
  \label{eq:sample mean}
\end{equation}
where $\hat{G}\left(\Omega_k\right)$ is the FRF estimate obtained from $\hat{\Theta}$ in the non-parametric identification part \eqref{eq:estimateFRF} and $\hat{T}(\Omega_k)$ is the transient term given as $\hat{T}\left(\Omega_k\right)=\hat{\Theta}_{\left[:, K(R+1)+1\right]}$ (the $K(R+1)+1$-th column of $\hat{\Theta}$). The sample covariance $\hat{C}_W(k)$ of $\hat{W}(k)$ can be calculated from the noise covariance model obtained from the non-parametric part as
\begin{equation}
  \hat{C}_W(k) = (Z_n^\Hn(Z_nZ_n^\Hn)^{-1}Z_n)_{[n+1,n+1]}\hat{C}_V(k).
  \label{eq:sampled covariance}
  \vspace{-5pt}
\end{equation}
 \begin{remark}
The asymptotic behaviors and the related proof of the sample mean and sample covariance are shown in \cite{pintelon2010estimation-2} and \cite[Chapter 12]{pintelon2012system}.\qed
\end{remark}

By replacing $W(k)$ in \eqref{eq:SK_iter_ori} and the frequency weighting $\hat{C}_V(k)$ with $\hat{W}(k)$ and $\hat{C}_W(k)$, respectively, we can formulate the frequency-domain input/output data criterion as
\begin{equation}
  \hat{\theta }=\arg\underset{\theta }{\mathop{\min }}\,\frac{1}{N}\sum\limits_{k=1}^{N}\left||M_1(k,\theta)|\right|^2_\Fn,
  \label{eq: i/o data criterion 2}
\end{equation}
with
\vspace{-20pt}
\begin{align}
    M_1 &=  \Bigl[ {\hat{C}_W(k)}^{\frac{1}{2}} A({{\Omega }_{k}}, \theta )^{i-1} \Bigl]^{-1}
    \Bigl[ A({{\Omega }_{k}}, \theta ) \hat{W}(k) \nonumber\\
    &\quad - B({{\Omega }_{k}}, \theta ) R(k) - C({{\Omega }_{k}}, \theta ) \Bigl].
\end{align}
Collecting all input/output data for all frequencies and the frequency weighting in the regression matrix $Q$, and the unknown parameters in the vector $\theta$ for each iteration,
\begin{equation}
    M_1 = Q\theta.
    \label{eq:MQ_theta}
\end{equation}
Due to the non-monicity of the polynomial matrix $A(p)$ and since the parameters of $A(p)$ and $B(p)$ might be partially known, we construct a constrained iterative weighted least squares (IWLS) optimization using SK-iteration, where the constraints include the known parameters and the interconnection structure. Based on criterion (\ref{eq: i/o data criterion 2}) and the constraints, the parametric estimation in each iteration is given by
\begin{equation}
\begin{aligned}
      \hat{\theta} = \arg\min_{\theta} \theta^\Hn Q^\Hn Q\theta, \\
  \text{subject to } \Gamma\theta=\upsilon,
  \label{parametric kkt}
\end{aligned}
\end{equation}
where the constraint $\Gamma\theta=\upsilon$ follows the Condition \ref{constraint_con} in Proposition \ref{condition:para}.

The Lagrangian multiplier $\lambda$ and Karush–Kuhn–Tucker (KKT) conditions can be used to solve this optimization problem \cite{chong2013optimization}, which gives
\begin{equation}\label{eq:kkt condition}
  \begin{aligned}
    &\mathcal{L}=\theta^\Hn Q^\Hn Q\theta + \lambda(\Gamma\theta-\upsilon),\\
    & \frac{\partial \mathcal{L}}{\partial \theta }=2 Q^\Hn Q \theta +{{\Gamma }^{\Tn}}\lambda =0, \\
    & \frac{\partial \mathcal{L}}{\partial \lambda }=\Gamma \theta -\upsilon =0.
  \end{aligned}
\end{equation}
The estimation of the parameters $\theta$ is given as
\begin{equation}
    \left[ \begin{matrix}
   {\hat{\theta}}  \\
   {\hat{\lambda} }  \\
\end{matrix} \right]={{\left[ \begin{matrix}
   2 Q^\Hn Q & {{\Gamma }^{\Tn}}  \\
   \Gamma  & \mathrm{O}  \\
\end{matrix}\right]}^{-1}}\left[ \begin{matrix}
   \mathrm{O}  \\
   \upsilon   \\
\end{matrix} \right],
\label{eq:estimation of parametric}
\end{equation}
where $\mathrm{O}$ is a zero-matrix with proper dimensions and $\hat{\lambda}$ are the estimated Lagrange multipliers.
\begin{remark}
The structures of the regression matrix $Q$, parameter vector $\theta$, selection matrix $\Gamma$, and constant vector $\upsilon$ are described in Appendix~\ref{App: The structure of Q1 theta1}. The parameterization of $A(p,\theta)$ is symmetric, which fixed the off-diagonal terms of $A(p,\theta)$ directly in the symmetric structure.\qed
\end{remark}

\subsection{Step 3: reduce bias with structured MLE estimation}\label{sec:MLE}
The structured polynomial matrices $\hat{A}(\hat{\theta})$ and $\hat{B}(\hat{\theta})$ have been identified from the last step. However, the convergence of the SK-iteration does not result in the exact minimum of the nonlinear cost criterion \eqref{eq:nonliear_criterion}, leading to a biased estimated model. To improve the accuracy of the estimated model, we use the result of the IWLS in \eqref{eq:estimation of parametric} as an initial estimate for the sample maximum likelihood estimator (SMLE). The SMLE is asymptotically unbiased and asymptotically efficient \cite{pintelon2012system}.
The SMLE is given in line with \cite{pintelon2010estimation-2} as
\begin{equation}\label{eq: SMLE}
    \hat{\theta }=\arg\underset{\theta }{\mathop{\min }}\,\frac{1}{N}\sum\limits_{k=1}^{N}\left||M_2(k,\theta)|\right|^2_F,
\end{equation}
with
\begin{align}
    &M_2= \hat{C}_W(k)^{-\frac{1}{2}}\left[\hat{W}(k)-G(\Omega_k,\theta)R(k)-T(\Omega_k,\theta) \right],\\
    &G(\Omega_k,\theta) = A({{\Omega }_{k}},\theta )^{-1}B({{\Omega }_{k}},\theta ),\\
    &T(\Omega_k,\theta) = A({{\Omega }_{k}},\theta )^{-1}C({{\Omega }_{k}},\theta ),
   \label{eq: i/o data criterion 3}
\end{align}
where $\hat{W}(k)$ and $\hat{C}_W(k)$ are the sample mean and covariance, obtained in \eqref{eq:sample mean} and \eqref{eq:sampled covariance}, respectively.

 Since we focus on the values of the components in the physical network, keeping the network structure during the SMLE is essential. The symmetric structure of $A(p)$ is incorporated in the parameterization and the known dynamic $B(p)$ is fixed. The SMLE cost is minimized using the solver 'lsqnonlin' in Matlab with the fixed parameters included and solved using \eqref{eq:estimation of parametric} as the initial estimate for the GN algorithm.

\subsection{Full network identification algorithm}
The steps given above describe the frequency-domain identification procedure, which can be summarized in the following algorithm.
\begin{Algorithm}\label{algorithm}
 \vspace{-5pt}
The frequency-domain identification algorithm for diffusively coupled linear networks is given as:
\begin{enumerate}
    \item Apply the LPM method to estimate the non-parametric $\hat{G}$ with \eqref{eq:LPM-LS} and the noise covariance $\hat{C}_V$ with \eqref{noise covariance}.
    \item Apply the input/output data criterion \eqref{eq: i/o data criterion 2} with sample mean \eqref{eq:sample mean} and covariance \eqref{eq:sampled covariance} leading to the IWLS with constraint \eqref{parametric kkt} to estimate the parameters $\hat{\theta}$ in a structured polynomial model, which is solved using the SK-iteration.
    \item Use the result \eqref{eq:estimation of parametric} as an initial estimate for the SMLE \eqref{eq: SMLE} to obtain asymptotically unbiased and efficient parameter estimates by an iterative GN algorithm.\qed
\end{enumerate}
\end{Algorithm}
\begin{remark}
    Achieving consistent estimates of DCNs using this algorithm requires satisfying all conditions specified in Proposition~\ref{condition:para} for network identifiability. It is also necessary that the true model is one of the candidate parametric models in the model set ($P_{0} \in \mathcal{M}$), and the system must be excited at sufficient frequencies for data informativity\cite{lizanidentifibility}.\qed
\end{remark}

\section{FREQUENCY-DOMAIN SUBNETWORK IDENTIFICATION FOR DIFFUSIVELY COUPLED NETWORKS}\label{sec:local id}
The full network identification algorithm for DCNs demands to measure or excite all nodes. However, when some nodes are inaccessible or users only want to identify part of the network, only part of the nodes in a DCN are measured. Thus, we present a frequency-domain subnetwork identification algorithm for DCNs to identify local dynamics with partial measurements in this section. This methodology is built on the full network identification Algorithm \ref{algorithm} and adapts the time-domain subnetwork identification algorithm from \cite{lizan2022} to frequency domain for estimating CT DCNs. In this section, $(1)$ the immersed network will be introduced to model the network with partial measurement ; $(2)$ the frequency-domain subnetwork identification procedure will be presented ; $(3)$ this procedure will be concluded in an algorithm.

\subsection{The immersed network}\label{sec: immersed network}
Identifying the subnetwork with partial measurement can save much instrument cost. In contrast to full network identification, where all nodes are measured, subnetwork identification requires measuring only a specific subset of node signals. Those unmeasured node signals can be eliminated from the model using Gaussian elimination, which is known as Kron reduction \cite{DorflerKron} or immersion \cite{dankerslocal}\cite{lizan2022}. Several algorithms have been developed to identify the local module or subnetwork of a dynamic network, but without taking into account the undirected physical network structure \cite{ramaswamy2022frequency}\cite{dankerslocal}. Here, following the literature \cite{lizan2022}, the immersed network of a DCN can be given as follows.

\begin{definition}(Immersed DCN)
    Consider a DCN model as defined in \eqref{def:diffusively network}, the node signals $w(t)$ are divided into two groups; one is the measurement groups $w_{\mathfrak{M}}(t)$ with the measurement set $ \mathfrak{M} = \left\{j|w_j \in w_{\mathfrak{M}} \right\}$, the other is the unmeasured immersion groups $w_{\mathfrak{I}}(t)$ with the immersion set $ \mathfrak{I} = \left\{j|w_j \notin w_{\mathfrak{M}} \right\}$. The immersed DCN is defined as
    \begin{equation}
        A_{im}(p) w_{\mathfrak{M}}(t)= B_{im}(p)r(t) + F_{im}(p)e_{\mathfrak{M}}(t),
        \label{eq: def-immersed dcn}
    \end{equation}
    where $A_{im}(p)$ and $B_{im}(p)$ are polynomial matrices. Since we are using the frequency-domain identification method, there is no need to parameterize the noise model. We can consider $F_{im}(p)e_{\mathfrak{M}}(t)$ as $V_{im}(t)$.\qed
\end{definition}

According to \cite{lizan2022},  to identify the target subnetwork, it is sufficient to measure at least the target subnetwork node signals and all neighbor node signals of the target subnetwork. Therefore, the measurement set can also be divided into two parts, the target subnetwork part $J = \{j|w_j \in w_J\}$ and the neighbors of the target subnetwork $D = \{j|w_j \in w_D\}$ with $J \cup D = \mathfrak{M}$. The partition of the original DCN in \eqref{eq:def-idmodel} can be expressed as:
\begin{equation}
    \underbrace{\left[\begin{matrix}
        {A}_{JJ} & {A}_{JD} & 0\\
        {A}_{DJ} & {A}_{DD} & {A}_{D\mathfrak{I}}\\
        0 & {A}_{\mathfrak{I}D} & {A}_{\mathfrak{I}\mathfrak{I}}\\
    \end{matrix}\right]}_{A(p)} \underbrace{\left[\begin{matrix}
        w_J(t) \\
        w_D(t) \\
        w_\mathfrak{I}(t)
    \end{matrix}\right]}_{w(t)} = \underbrace{\left[\begin{matrix}
        {B}_J \\
        {B}_D \\
        {B}_\mathfrak{I}
    \end{matrix}\right]}_{B(p)}r(t) + \underbrace{\left[\begin{matrix}
        {V}_J(t) \\
        {V}_D(t) \\
        {V}_\mathfrak{I}(t)
    \end{matrix}\right]}_{V(t)},
    \label{eq:partition original network}
\end{equation}
where,
\begin{equation}
\begin{aligned}
     &A_{\mathfrak{M}\mathfrak{M}} =\left[\begin{matrix}
        {A}_{JJ} & {A}_{JD} \\
        {A}_{DJ} & {A}_{DD}
    \end{matrix}\right],
        A_{\mathfrak{M}\mathfrak{I}} =\left[\begin{matrix}
        0 \\
        {A}_{D\mathfrak{I}}
    \end{matrix}\right],
        A_{\mathfrak{I}\mathfrak{M}} =\left[\begin{matrix}
        0 & {A}_{D\mathfrak{I}}
    \end{matrix}\right], \\
        &B_{\mathfrak{M}} =\left[\begin{matrix}
        {B}_{J}  \\
        {B}_{D}
    \end{matrix}\right],
        V_{\mathfrak{M}} =\left[\begin{matrix}
        {V}_{J}  \\
        {V}_{D}
    \end{matrix}\right].
\end{aligned}
\end{equation}
By eliminating the unmeasured node signals $w_\mathfrak{I}(t)$ in \eqref{eq:partition original network}, the partition result of the immersed DCN in \eqref{eq: def-immersed dcn} is
\begin{equation}
    \underbrace{\left[\begin{matrix}
        \Bar{A}_{JJ} & \Bar{A}_{JD} \\
        \Bar{A}_{DJ} & \Bar{A}_{DD}
    \end{matrix}\right]}_{A_{im}(p)} \underbrace{\left[\begin{matrix}
        w_J(t) \\
        w_D(t)
    \end{matrix}\right]}_{w_{\mathfrak{M}(t)}} = \underbrace{\left[\begin{matrix}
        \Bar{B}_J \\
        \Bar{B}_D
    \end{matrix}\right]}_{B_{im}(p)}r(t) + \underbrace{\left[\begin{matrix}
        \Bar{V}_J(t) \\
        \Bar{V}_D(t)
    \end{matrix}\right]}_{V_{im}(t)},
    \label{eq:partition immersed network}
\end{equation}
with
    \begin{equation}
    \begin{aligned}
        \Bar{A}_{JJ} &= d_{\mathfrak{II}}A_{JJ}, \quad  \Bar{A}_{JD} = d_{\mathfrak{II}}A_{JD},& \\
        \Bar{A}_{DJ} &= d_{\mathfrak{II}}A_{DJ}, \quad  \Bar{B}_{J}  = d_{\mathfrak{II}}B_{J},&
    \end{aligned}
    \label{eq:relation Aim and Ao}
    \end{equation}
where $A_{JJ}(p)$ contains the dynamics of the target subnetwork of the original model $A(p)$, $A_{JD}(p) = A_{DJ}(p)^\Tn$ contains the dynamics that in the interconnection between the target subnetwork and the neighbor nodes; $B_{J}(p)$ contains the dynamics of the excitation signal that enters the target subnetwork; $d_{\mathfrak{II}}(p)$ is a scalar polynomial to ensure that the matrices for the immersed network model stay in polynomial during the Gaussian elimination process.
\begin{remark}
    $d_{\mathfrak{II}}(p)$ is set to ${\mathrm{det}(A_{\mathfrak{I}\mathfrak{I}})}$ if ${\mathrm{det}(A_{\mathfrak{I}\mathfrak{I}})}$ and ${\mathrm{adj}(A_{\mathfrak{I}\mathfrak{I}})}$ have no common factors ($\mathrm{det}({A_{\mathfrak{I}\mathfrak{I}}})$ and ${\mathrm{adj}(A_{\mathfrak{I}\mathfrak{I}})}$ is the determinant and the adjugate of the polynomial matrix ${A_{\mathfrak{I}\mathfrak{I}}}$, respectively). In case a common factor is present, dividing by the greatest common divisor of these elements is required to achieve a uniquely represented immersed DCN. The proof of invariant local dynamics is shown in \cite{lizan2022}. \qed
\end{remark}

\subsection{The subnetwork identification procedure}\label{sec: subnetwork procedure}
Identifying DCNs in the frequency domain involves three main steps. $(1)$ The first step follows Section~\ref{sec:non-parameteric}, to estimate the FRF and noise covariance of the immersed DCN. $(2)$ The second step involves using them to estimate the parametric immersed DCN model with user-defined constraints, following the same procedure as in Sections~\ref{sec:parametric id} and~\ref{sec:MLE}. Since the scalar $d_{\mathfrak{II}}$ is unknown, at least one constraint on a non-zero parameter is needed to guarantee a unique solution, which leads to a (scaled) immersed DCN estimate compared to the actual immersed DCN. $(3)$ The third step is to recover the scaled parameters of the target subnetwork by giving at least one known parameter of the original (not immersed) DCN, such as the parameter of the excitation matrix $B_{\mathfrak{M}}(p)$, in line with \cite{lizan2022}.

To employ the full network identification algorithm in the first two steps, some conditions similar to Proposition~\ref{condition:para} must be satisfied. Consider a parametric model set $\mathcal{M}_{im}$ of the immersed DCN
\begin{equation}
\begin{aligned}
    \mathcal{M}_{im} = \left\{A_{im}(p,\eta),B_{im}(p,\eta),C_{im}(p,\eta)\right\},
\end{aligned}
\end{equation}
where $\eta$ collects all the unknown parameters of matrices $A_{im}(p),B_{im}(p)$ and the transient $C_{im}(p)$. The data generated immersed DCN is denoted by $\mathcal{S}_{im} = \left\{A_{im}^0(p),B_{im}^0(p),C_{im}^0(p),F_{im}^0(p)\right\}$. We only need $P_{{im}_{0}} \in \mathcal{M}_{im}$ for a consistent estimate, with $P_{{im}_{0}} = \left(A_{im}^0(p,\eta^0),B_{im}^0(p,\eta^0),C_{im}^0(p,\eta^0)\right)$.

\begin{Proposition}
    The estimated parameters $\hat{\eta}$ give a consistent estimate of the (scaled) immersed network if the following conditions hold \cite{lizan2022}.
    \begin{enumerate}
        \item The polynomials $A_{im}(p,\eta)$ and $B_{im}(p,\eta)$ are left-coprime.
        \item There exists a permutation matrix $P$ that leads to $\left[
             A_{im,0}\text{ }A_{im,1}\text{ }\cdots\text{ }A_{im,n_{aim}}\text{ }B_{im,0}\text{ }\cdots\text{ }B_{im,n_{bim}}
        \right]P =\left[D(\eta) \text{ }U(\eta)\right]$ with $D(\eta)$ square, diagonal, and full rank.
        \item There exists at least one excitation signal, $K\geq1$.
        \item There exists at least one constraint on the parameters of $A_{im}(p,\eta)$ and $B_{im}(p,\eta)$ that ensures $\Gamma \eta = \upsilon \neq 0$, where $\Gamma$ is a full-row rank matrix. \label{cond: immersed constraint}
        \item The system is excited at sufficient frequencies. \qed
    \end{enumerate}
    \label{proposition: immersion}
\end{Proposition}
\begin{remark}
    The proof of Proposition \ref{proposition: immersion} is given in \cite{lizan2022}. Notice that the constraint in Condition \ref{cond: immersed constraint} is user-defined, which could lead to a scaled immersed DCN.
    \qed
\end{remark}

Following the frequency-domain full network identification procedure shown in Section~\ref{Sec:fdomain identificaiton algorithm}, the (scaled) immersed DCN $\hat{A}_{im}(\hat{\eta})$ and $\hat{B}_{im}(\hat{\eta})$ can be estimated. Then, the target subnetwork can be identified from this estimated (scaled) immersed DCN. From \eqref{eq:partition immersed network} and \eqref{eq:relation Aim and Ao}, the estimated (scaled) immersed DCN has the following relation with the target subnetwork,
\begin{equation}
\begin{aligned}
    \hat{A}_{JJ}(\hat{\eta}) &= \alpha d_{\mathfrak{II}}{A}_{JJ},
    \hat{A}_{JD}(\hat{\eta}) = \alpha d_{\mathfrak{II}}{A}_{JD}, \\
    \hat{B}_{J}(\hat{\eta})  &= \alpha d_{\mathfrak{II}}{B}_{J},
    \label{eq:immerback}
\end{aligned}
\end{equation}
where $\alpha$ is the unknown scaling factor that $\hat{A}_{JJ}(\hat{\eta}) = \alpha \Bar{A}_{JJ}$, $\hat{A}_{JD}(\hat{\eta}) = \alpha \Bar{A}_{JD}$ and $\hat{B}_{J}(\hat{\eta}) = \alpha \Bar{B}_{J}$. Substitute the first relation ${A}_{JJ}^{-1}\hat{A}_{JJ} = \alpha d_{\mathfrak{II}}$ into the rest two equation of (\ref{eq:immerback}),
\begin{equation}
\begin{aligned}
       A_{JJ}\hat{A}_{JD}(\hat{\eta}) - \hat{A}_{JJ}(\hat{\eta}) A_{JD} & = M_3 = 0, \\
       A_{JJ}\hat{B}_{J}(\hat{\eta})   - \hat{A}_{JJ}(\hat{\eta}) B_{J} & = M_4 = 0.
       \label{eq:immerback2}
\end{aligned}
\end{equation}
Collecting all the frequency data of the known estimated part $\hat{A}_{JJ}(\Omega_k,\hat{\eta})$,$\hat{A}_{JD}(\Omega_k,\hat{\eta})$ and $\hat{B}_{J}(\Omega_k,\hat{\eta})$ in $Q_1(\hat{\eta})$, and all the unknown parameters of the target subnetwork $A_{JJ}(\theta_1)$, $A_{JD}(\theta_1)$ and $B_{J}(\theta_1)$ in parameter vector $\theta_1$, we have
\begin{equation}
    Q_1(\hat{\eta})\theta_1 = 0
\end{equation}
  Since $\hat{A}_{im}(\hat{\eta})$ and $\hat{B}_{im}(\hat{\eta})$ are already estimated as known parts, the estimation problem leads to a convex optimization problem. Employing the same idea as in \eqref{eq:kkt condition}, use the Lagrangian and KKT conditions to solve the least squares optimization with constraint,
\begin{equation}
    \begin{aligned}
        \hat{\theta}_1 = \arg\min_{\theta_1} \theta_1^\Hn Q_1^\Hn(\hat{\eta})Q_1(\hat{\eta})\theta_1, \\
        \text{subject to} \text{ } \Gamma_1\theta_1=\upsilon_1,
        \label{eq:parametric kkt immersed back}
    \end{aligned}
\end{equation}
where the constraint is set to guarantee a consistent estimate of the target subnetwork by giving at least one known nonzero polynomial coefficient $a_{ij}(p)$ or $b_{ij}(p)$ in $A_{JJ}(\theta_1)$, $A_{JD}(\theta_1)$ or $B_{J}(\theta_1)$, respectively.

\begin{remark}
    Setting the constraints $\Gamma_1\theta_1=\upsilon_1$ for recovering the target subnetwork from the estimated immersed DCN is similar to the previous full network case. But in this case, we follow the full parameterization structure as shown in \cite{Liangmsc}, by which we estimate all the parameters in $A_{JJ}(\theta_1), A_{JD}(\theta_1)$ and $B_{J}(\theta_1)$. This parameterization is easier to implement for the non-square matrix $[A_{JJ}(\theta_1)\text{ }A_{JD}(\theta_1)]$. \qed
\end{remark}

Since we chose the full parameterization structure for subnetwork identification, setting the constraint in \eqref{eq:parametric kkt immersed back} to keep DCN structure is essential, i.e., the symmetric structure of $A_{JJ}(\theta_1)$,
\begin{equation}
    {a}_{{ij}}(p) - {a}_{{ji}}(p) = 0 \text{ with } i,j \in J.
\end{equation}
The constraint selection matrix $\Gamma_1$ and the constant vector $\upsilon_1$ are given similarly as $\Gamma$ and $\upsilon$ presented in Appendix~\ref{lagrangian constraint}, but the parameters are collected in a different sequence $\theta_1$ as shown in Appendix~\ref{App: The structure of Q3 theta3}. Thus, the approach involves applying a similar constraint method, but adapting the constraint settings to match the parameter sequence in $\theta_1$.

The estimation of the parameters $\theta_1$ are given as
\begin{equation}
    \left[ \begin{matrix}
   {\hat{\theta}_1}  \\
   {\hat{\lambda}_1 }  \\
\end{matrix} \right]={{\left[ \begin{matrix}
   2 Q_1(\hat{\eta})^\Hn Q_1(\hat{\eta}) & {{\Gamma_1 }^{\Tn}}  \\
   \Gamma_1  & \mathrm{O}  \\
\end{matrix}\right]}^{-1}}\left[ \begin{matrix}
   \mathrm{O}  \\
   \upsilon_1   \\
\end{matrix} \right],
\label{eq:estimation of parametric sub}
\end{equation}
where $\mathrm{O}$ are the zero matrices with the proper dimensions, and $\hat{\lambda}_1$ are the estimated Lagrange multipliers. The structure of the regression matrix $Q_1$ and the parameter vector $\theta_1$ are given in the Appendix~\ref{App: The structure of Q3 theta3}.

\subsection{Subnetwork identification algorithm} \label{sec: subnetwork procedure flow}
\begin{Algorithm}\label{subnetwork algorithm}
The frequency-domain identification algorithm for the diffusively coupled linear subnetwork is given:
\begin{enumerate}
    \item Apply the LPM method to estimate the non-parametric $\hat{\Theta}$ with \eqref{eq:LPM-LS} and noise covariance $\hat{C}_{V}$ with \eqref{noise covariance} of the immersed DCN \eqref{eq:partition immersed network}.
    \item Apply the input/output data criterion \eqref{eq: i/o data criterion 2} with sample mean \eqref{eq:sample mean} and covariance \eqref{eq:sampled covariance} leading to the IWLS with constraint \eqref{parametric kkt} to estimate the parameters $\hat{\eta}$ of the (scaled) immersed DCN \eqref{eq:partition immersed network} in a structured polynomial model, which is solved using the SK-iteration.
    \item Use the IWLS result \eqref{eq:estimation of parametric} as an initial estimate for the SMLE in \eqref{eq: i/o data criterion 3}. The asymptotically unbiased and efficient estimates \(\hat{A}_{im}(\hat{\eta})\), \(\hat{B}_{im}(\hat{\eta})\) of the immersed DCN \eqref{eq:partition immersed network} are obtained using the iterative GN algorithm.
    \item Apply the relation \eqref{eq:immerback2} to estimate the target subnetwork parameters $\hat{\theta}_1$ by optimizing the least squares with the constraint \eqref{eq:parametric kkt immersed back}, results the target subnetwork estimates \eqref{eq:estimation of parametric sub}.\label{sub: step4}     \qed
\end{enumerate}
\end{Algorithm}
\begin{remark}
    Note that the constraint of network topology and the known dynamics in step~\ref{sub: step4} are obtained based on the original (not immersed) network in \eqref{def:diffusively network}, because the immersion will not change the topology of the target subnetwork and the dynamics in $A_{JJ}, A_{JD}$, and $B_{J}$ of target subnetwork are invariant during the immersion. Although topology information is not necessary for identification, it provides information on the locations of zero terms in $A(p)$ and $A_{im}(p)$. Incorporating this into the constraints can enhance convergence speed and reduce computational complexity in \eqref{parametric kkt}, \eqref{eq: SMLE}, and \eqref{eq:parametric kkt immersed back}. Moreover, it improves the precision of the estimation by reducing the degrees of freedom in the parameter space, which can reduce bias and variance of the estimates. In addition, this algorithm is fully in the frequency domain to identify the frequency-domain model, resulting in a uniform identification procedure for both CT and DT models. \qed
\end{remark}

\section{SIMULATION EXPERIMENTS}\label{sec:FDD}
Several experiments are performed in MATLAB simulation to test the frequency-domain identification algorithm for estimating physical component values in CT DCNs. We will show the consistent parameter estimation and In-Circuit Testing applications for full network identification and subnetwork identification.

\subsection{Full network identification experiments}
The following simulation examples serve to illustrate that the component values of a full DCN can be consistently estimated from only a single excitation signal with full node measurements and to show an In-Circuit Testing application using Algorithm~\ref{algorithm}.

Consider the 10-node RLC network as shown in Fig.~\ref{fig:RLC circuit 10nodes}, where each $j$-th node, $j=1,\ldots,10$, has the following measurement components connected to the ground node: $C_{j0}=2$ $\mu$F, $R_{j0}=500$ $\Omega$, and $L_{j0}=18$ mH. The coefficients of the components in the interconnections between the nodes are given as ${\theta}_{hcomp}$ in Table~\ref{tab:10nodesfull_defect}.

This RLC circuit can be expressed as a second-order CT DCN model, the voltage $U(t)$ is considered as the output node signal and the current $I(t)$ is treated as the excitation   signal,
\begin{equation}
    \begin{aligned}
        &{{C}_{j0}}{{\ddot{U}}_{j}(t)}+\sum_{k\in \mathcal{N}_j}{{C}_{jk}}\left( {{{\ddot{U}}}_{j}(t)}-{{{\ddot{U}}}_{k}(t)} \right)+\frac{1}{{{R}_{j0}}}{{\dot{U}}_{j}(t)} \\
        & +\sum_{k\in \mathcal{N}_j}\frac{1}{{{R}_{jk}}}\left( {{{\dot{U}}}_{j}(t)}-{{{\dot{U}}}_{k}(t)} \right)+\frac{1}{{{L}_{j0}}}{{U}_{j}} \\
        &+\sum_{k\in \mathcal{N}_j}\frac{1}{{{L}_{jk}}}\left( {{U}_{j}(t)}-{{U}_{k}(t)} \right)={{\dot{I}}_{j}(t)},
        \label{eq:RLC_2nd_diff}
    \end{aligned}
\end{equation}
where $U_j(t),j = 1,2,...,\text{N}$ are the N interconnected node signals; $L_{jk}\ge0, R_{jk}\ge0,{C}_{jk}\ge0, L_{jj} = 0, R_{jj}=0, C_{jj}=0$ are the real-valued coefficients of the whole coefficient matrix $L$, $R$ and $C$; $L_{jk} = L_{kj}$ , $R_{jk}=R_{kj}$ and $C_{jk}=C_{kj}$ are the diffusively coupled constraints; $\mathcal{N}_j$ is the set of indexes of the node signals $U_k(t)$ with connection to $U_j(t)$, $k\ne j$; $I_j(t)$ are the external input signals and ${{\dot{I}}_{j}(t)}$ are the first order derivative of the input signals; $\ddot{U}_j(t)$ and $\dot{U}_j(t)$ are the second order and the first order derivative of the node signals $U_j(t)$ with respect to time $t$, respectively. Collecting all the coefficients into polynomial matrices as \eqref{def:diffusively network} lead to a DCN model, where the symmetric $A(p)$ parameters are given as $a_{jj,0} = \frac{1}{L_{j0}}$, $a_{jj,1} = \frac{1}{R_{j0}}$, and $a_{jj,2} = {C_{j0}}$; and $a_{jk,0} = -\frac{1}{L_{jk}}$, $a_{jk,1} = -\frac{1}{R_{jk}}$, and $a_{jk,2} = -{C_{jk}}$ for $j\neq k$. The rest of the parameters in $A(p)$ are 0 for the absent connections. The parameters of $B(p)$ are given as $b_{jk,1} = 1$ for the $k$-th excitation entering the $j$-th node, and the rest of the parameters in $B(p)$ are 0. Thus, the parametric model orders are defined as $n_a = 2$, $n_b = 1$, and $n_c = 1$.
\subsubsection{Consistent parameter estimation}
The excitation signal $r(t)$ is an independent zero-mean white noise with variance $\sigma_r^2 = 1$ entering only at node 3. The noise signal $e(t)$ is a normal distributed zero-mean white noise with variance $\sigma_e^2 = 1$ entering all nodes. The sampling frequency is set at 20 kHz to cover all the dynamics of the components, and the identification frequency band is set between $f_{min} = 500$ Hz and $f_{max} = 4$ kHz. The constraints incorporate the known input matrix $B(p)$ and the known topology information indicated in the matrix $A(p)$.

To show that the parameters can be consistently estimated with a single excitation, we generated a set of experiments with different data lengths $N$. The choice of $N$ and the corresponding set number are shown in Table~\ref{tab:datalength}. Each set of experiments includes 100 Monte Carlo (MC) runs with independent excitation and noise signals.
\begin{table}[tbp]
\vspace{-10pt}
\caption{Experiment set numbers with corresponding data lengths $N$.}
\vspace{-10pt}
\begin{center}
\begin{tabular}{|c|c|c|c|c|c|c|}
\hline
\textbf{Set} & 1 & 2 &3&4&5\\
\hline
$N$ & $10^3$ & $2\times10^3$ &$4\times10^3$ & $8\times10^3$ & $16\times10^3$   \\
\hline
\textbf{Set} &6&7&8&9&10 \\
\hline
$N$& $32\times10^3$ &$64\times10^3$ & $128\times10^3$ & $256\times10^3$ &$512\times10^3$\\
\hline
\end{tabular}
\label{tab:datalength}
\end{center}
\end{table}
\begin{figure}[tbp]
\vspace{-10pt}
    \centering
\includegraphics{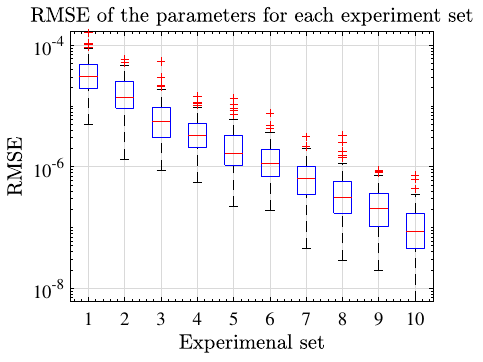}
    \vspace{-10pt}
    \caption{Boxplot of the RMSE of the coefficients of the components for each experimental set (full network identification).}
    \label{fig:CT_10node_consist}
    \vspace{-10pt}
\end{figure}

The box plots of the relative mean squared error (RMSE) of the coefficients of the components for each experimental set are shown in Fig.~\ref{fig:CT_10node_consist}. The relative mean squared error of the coefficients of the components is given as
\begin{equation}
  \mathrm{RMSE} = \frac{\|\hat{\theta}_{comp}-{\theta}_{comp}^0 \|_2^2}{\|{\theta}_{comp}^0\|_2^2},
\end{equation}
where $\hat{\theta}_{comp}$ and ${\theta}_{comp}^0$ collect the estimated and actual component values, respectively. The $\hat{\theta}_{comp}$ is uniquely recovered from the estimation of the DCN parameters $\hat{\theta}$. In Fig.~\ref{fig:CT_10node_consist}, it can be seen that RMSE decreases as $N$ increases, which supports the achievement of consistent identification. When the data length $N$ extends to infinity, the estimated parameters converge to the actual parameters.
\subsubsection{Fault detection and diagnosis}
The ability to detect and diagnose faults for multiple open circuits and dynamic changes that occur simultaneously in this 10-node RLC network is also shown. Consider changed dynamics in $R_{13}$, $R_{36}$, $R_{89}$, and $L_{25}$ and open circuit in $R_{45}$ and $L_{56}$, leading to new actual values indicated by ${\theta}_{comp}^0$ in Table~\ref{tab:10nodesfull_defect}. We keep the same excitation situation as before and increase the noise power to $\sigma_e^2 = 100$ at all nodes. The data length is $N=20000$.

\begin{figure}[tbp]
    \centering
    \includegraphics[width=0.85\columnwidth]
{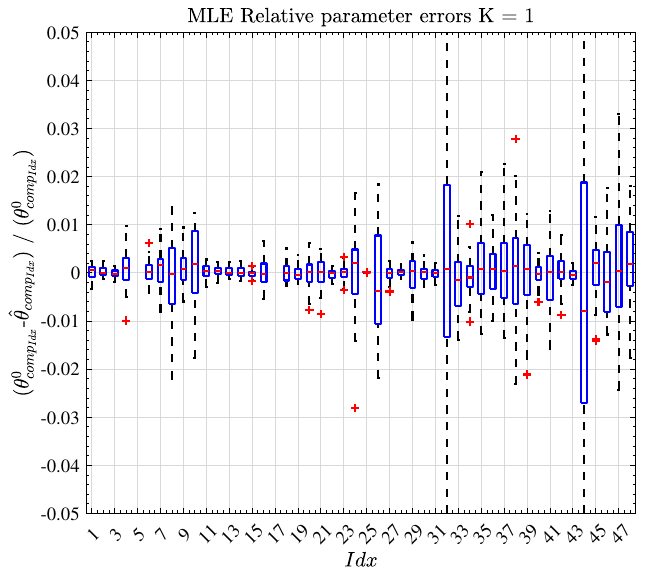}
    \vspace{-15pt}
    \caption{Relative parameters errors for the 10-node defect model. }
    \label{fig:CT_10node_defect}
\end{figure}

The relative parameter errors (RPEs) of the estimation results are shown in Fig.~\ref{fig:CT_10node_defect}. For each ungrounded component, the actual coefficient value (${\theta}_{comp}^0$) and the mean estimated coefficient value for 50 MC runs (CT est) are shown in Table~\ref{tab:10nodesfull_defect}. It can be seen that all mean estimated values, including the ones for the faulty components, are very close to the actual values. Moreover, all RPEs stay between $\pm 5\%$ for the healthy components as well as for the faulty components ($R_{13}$, $R_{36}$, $R_{45}$, $R_{89}$, $L_{25}$, and $L_{56}$), as shown in Fig.~\ref{fig:CT_10node_defect}. Notice that components ${Idx}$ 19 to 48, which are the grounded components (arranged in order of $C_{j0}$, $ R_{j0}$, $L_{j0}$, $j=1,\ldots,10$) show a higher variance compared to the others. This occurs as the interconnected components are represented twice in the model, resulting in the use of more data to estimate these components than is used for the grounded components. This experiment shows that all components of a physical network are correctly identified. By comparing the identified results with the ideal healthy components, we can detect the physical value changes of the components and diagnose the faults. This shows that the algorithm can be applied to FDD in complex physical networks.

\begin{table}[htbp]
    \caption{10-node defect network component values.}
    \vspace{-15pt}
    \begin{center}
    \scalebox{0.95}{
    \begin{tabular}{|c|c|c|c|c|c|}
        \hline
        $\boldsymbol{Idx}$ & \textbf{Comp} & $\boldsymbol{{\theta}_{hcomp}}$ & $\boldsymbol{{\theta}_{comp}^0}$ & \textbf{CT est}  & \textbf{Unit} \\
        \hline
        \textcolor{red}{1}&\textcolor{red}{\(R_{13}\)} & \textcolor{red}{100} & \textcolor{red}{200} & \textcolor{red}{200.06} &\textcolor{red}{\(\Omega\)} \\
        2&\(R_{23}\) & 200 & 200 & 200.04  &\(\Omega\)\\
        3&\(R_{34}\) & 150 & 150 & 149.98  &\(\Omega\)\\
        \textcolor{red}{4}&\textcolor{red}{\(R_{36}\)} & \textcolor{red}{180} & \textcolor{red}{500} & \textcolor{red}{500.40} &\textcolor{red}{\(\Omega\)} \\
        \textcolor{red}{5}&\textcolor{red}{\(R_{45}\)} & \textcolor{red}{350} & \textcolor{red}{Inf} & \textcolor{red}{Inf} &\textcolor{red}{\(\Omega\)} \\
        6&\(R_{38}\) &  180 & 180 & 180.37  &\(\Omega\) \\
        7&\(R_{56}\) & 160 & 160 & 160.15 &\(\Omega\)\\
        {8}&{\(R_{57}\)}  & 120 & {120} & {119.93} &{\(\Omega\)}\\
        \textcolor{red}{9}&\textcolor{red}{\(R_{89}\)} & \textcolor{red}{160} & \textcolor{red}{500} & \textcolor{red}{500.49} &\textcolor{red}{\(\Omega\)}\\
        10&\(R_{910}\) & 120& 120 & 120.13 &\(\Omega\)\\
        11&\(L_{19}\) & 5 & 5  &  5.00 & mH \\
        12&\(L_{29}\) & 3& 3 &  3.00   & mH \\
        13&\(L_{23}\) & 10& 10 &  10.00   & mH \\
        \textcolor{red}{14}&\textcolor{red}{\(L_{25}\)} & \textcolor{red}{15} & \textcolor{red}{1} &  \textcolor{red}{1.00}  & \textcolor{red}{mH} \\
        {15}&{\(L_{34}\)} & 12 & 12 & 12.00 & {mH} \\
        16&\(L_{45}\) & 20& 20 &  20.00 & mH \\
        \textcolor{red}{17}&\textcolor{red}{\(L_{56}\)} & \textcolor{red}{13} & \textcolor{red}{Inf} &  \textcolor{red}{Inf} & \textcolor{red}{mH} \\
        18&\(L_{89}\) &13  & 13 &  13.00  & mH \\
        \hline
    \end{tabular}}
    \label{tab:10nodesfull_defect}
    \end{center}
    \vspace{-25pt}
\end{table}

\subsection{Subnetwork identification experiments}
The following simulation examples serve to illustrate that the component values of a subnetwork in a DCN can be consistently estimated from only a single excitation signal with partial measurements and to show an In-Circuit Testing application using Algorithm~\ref{subnetwork algorithm}.

\begin{figure}[tbp]
\vspace{-15pt}
    \centering
    \includegraphics[width=0.45\textwidth]{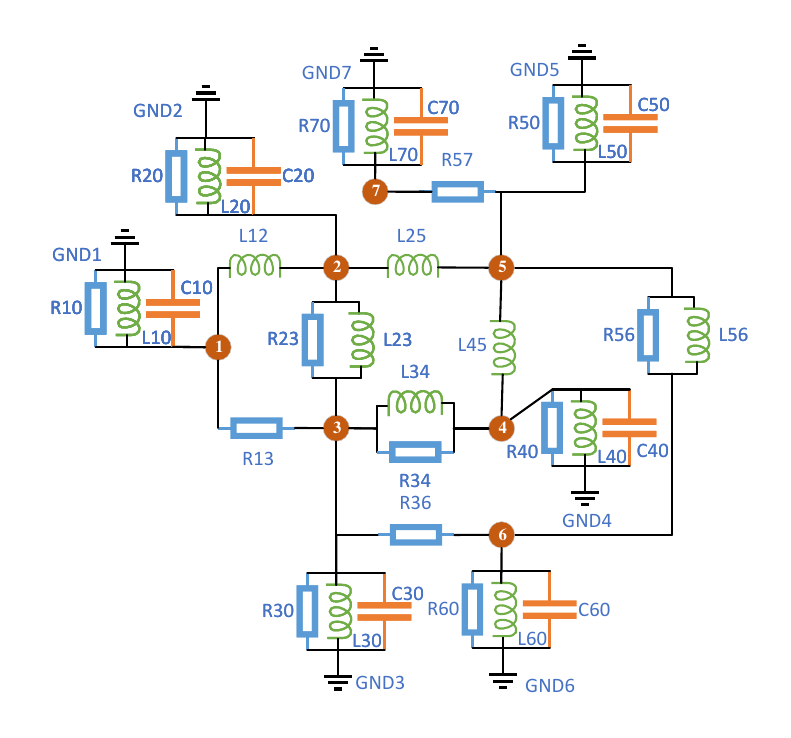}
    \vspace{-20pt}
    \caption{A seven-node RLC network circuit with inductors ($L_{jk}$), resistors ($R_{jk}$), capacitors ($C_{jk}$) and ground nodes($GND_{j}$).}
    \label{fig:RLC circuit 7nodes}
\end{figure}

Consider a 7-node RLC circuit as shown in Fig.~\ref{fig:RLC circuit 7nodes}, where each node has the following measurement components connected to the ground node: $C_{j0}=2$ $\mu$F, $R_{j0}=500$ $\Omega$, and $L_{j0}=18$ mH. The coefficients of the components in the interconnections between nodes are given in Table~\ref{tab:7nodesfull}.
Suppose that the target subnetwork we want to identify in Fig.~\ref{fig:RLC circuit 7nodes} is the interconnections between node 1 and node 2, and the grounded components connected to node 1 and 2. We measured the target subnetwork nodes 1 and 2 and their neighbor nodes 3 and 5. Given the excitation signal $r(t)$ as the independent zero-mean white noise with variance $\sigma_r^2 = 1$ entering only node 1, and the normal distributed zero-mean white noise as the noise signal $e(t)$ with variance $\sigma_e^2 = 1$ entering all 7 nodes. The sampling frequency is set at 20 kHz to cover all the dynamics of the components, and the identification frequency band is set between $f_{min} = 500$ Hz and $f_{max} = 6$ kHz. Furthermore,  the orders of the (scaled) immersed network model are defined as $n_{a_{im}} = 3$ of $A_{im}(p,\eta)$, $n_{b_{im}} = 2$ of $B_{im}(p,\eta)$, and $n_{c_{im}} = 1$ for the transient model. The orders of the parametric model of the target subnetwork are defined as $n_a = 2$, $n_b=1$.

\begin{table}[tbp]
\caption{Seven-node Network Component Values}
\begin{center}
\begin{tabular}{|c|c|c||c|c|c|}
\hline
\textbf{Resistor} & \textbf{Value} & \textbf{Unit} & \textbf{Inductor} & \textbf{Value} & \textbf{Unit} \\
\hline
\(R_{13}\) & 100 & \(\Omega\) & \(L_{12}\) & 5 & mH \\
\(R_{23}\) & 200 & \(\Omega\) & \(L_{23}\) & 10 & mH \\
\(R_{34}\) & 150 & \(\Omega\) & \(L_{25}\) & 15 & mH \\
\(R_{36}\) & 180 & \(\Omega\) & \(L_{34}\) & 12 & mH \\
\(R_{56}\) & 160 & \(\Omega\) & \(L_{45}\) & 20 & mH \\
\(R_{57}\) & 120 & \(\Omega\) & \(L_{56}\) & 13 & mH \\
\hline
\end{tabular}
\label{tab:7nodesfull}
\end{center}
\vspace{-10pt}
\end{table}

\subsubsection{Consistent parameter estimation}
\begin{figure}[tbp]
\vspace{-10pt}
    \centering
    \includegraphics{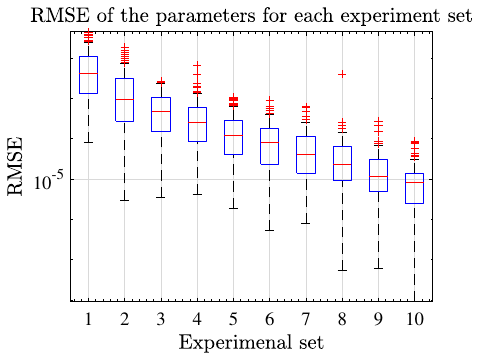}
    \vspace{-10pt}
    \caption{Boxplot of the RMSE of the coefficients of the components for each experimental set (subnetwork identification).}
    \label{fig:CT_7node_consist_imm}
    \vspace{-10pt}
\end{figure}

We generated a different set of experiments with a difference in the length of the data $N$. The choice of $N$ and the corresponding set are shown in Table~\ref{tab:datalength}. Each set of experiments includes 100 MC runs with independent excitation and noise signals. The results are shown in Fig.~\ref{fig:CT_7node_consist_imm}. It can be seen from this figure, RMSE decreases as N increases, supporting a consistent identification. When the data length $N$ is extended to infinity, the estimated parameters converge to the actual parameters.

\subsubsection{Fault detection and diagnosis}
\begin{figure}[htbp]
    \centering
    \includegraphics[width=0.95\columnwidth]{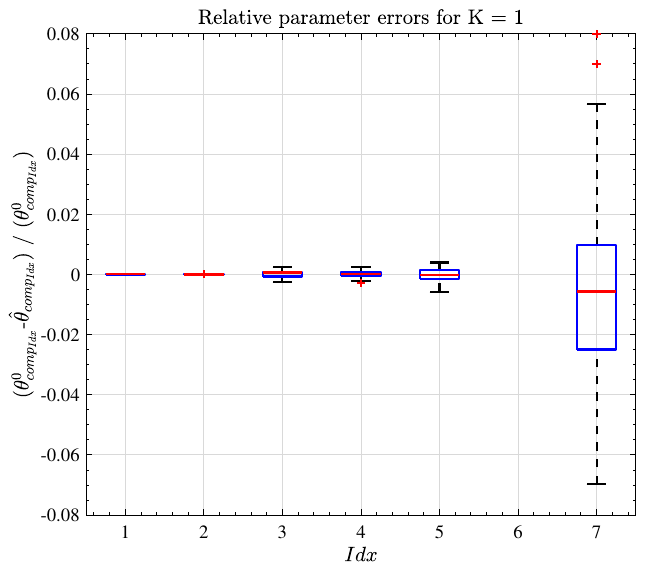}
    \vspace{-10pt}
    \caption{Relative parameters errors for the target subnetwork of the 7-node defect model. }
    \label{fig:CT_7node_defect_imm}
\end{figure}

 The ability to detect and diagnose faults for multiple open circuits and dynamic changes that occur simultaneously in the target subnetwork of this 7-node RLC network is shown. The data length is set as $N=40000$. Defects are given as changed dynamics in $R_{10}$ and $L_{12}$, open circuit in $R_{20}$, leading to new actual values indicated by ${\theta}_{comp}^0$ in Table~\ref{tab:7nodesimmersed_defect}. The RPEs of the estimation results for the CT model are shown in Fig.~\ref{fig:CT_7node_defect_imm}. The actual coefficient value of each component  (${\theta}_{comp}^0$) and the mean estimated parameter values of 50 MC runs (CT est) are shown in Table~\ref{tab:7nodesimmersed_defect}.

\begin{table}[tbp]
    \caption{Seven-node immersed defect Network Component Values}
    \begin{center}
    \hspace*{-5mm} 
    \begin{tabular}{|c|c|c|c|c|c|c|c|}
    \hline
    $\boldsymbol{Idx}$ & \textbf{Comp} & $\boldsymbol{{\theta}_{hcomp}}$ & $\boldsymbol{{\theta}_{comp}^0}$ & \textbf{CT est}  & \textbf{Unit} \\
    \hline
    \textcolor{red}{1}&\textcolor{red}{\(L_{12}\)} & \textcolor{red}{5}& \textcolor{red}{10}  &  \textcolor{red}{10.00}  & \textcolor{red}{mH} \\
    2&\(C_{10}\) & 2 & 2  & 2.00  &  \(\mu\)F \\
    \textcolor{red}{3}&\textcolor{red}{\(R_{10}\)} & \textcolor{red}{500}& \textcolor{red}{200} & \textcolor{red}{200.04}  &\textcolor{red}{\(\Omega\)}\\
    4&\(L_{10}\) & 18 & 18  & 18.00 & mH \\
    5&\(C_{20}\) & 2 & 2  & 2.00 & \(\mu\)F \\
    \textcolor{red}{6}&\textcolor{red}{\(R_{20}\)} & \textcolor{red}{500}& \textcolor{red}{Inf} & \textcolor{red}{4.600e+5} &\textcolor{red}{\(\Omega\)}\\
    7&\(L_{20}\) & 18 & 18  & 17.92 & mH \\
    \hline
    \end{tabular}
    \label{tab:7nodesimmersed_defect}
    \end{center}
    \vspace{-10pt}
 \end{table}

 Fig.~\ref{fig:CT_7node_defect_imm} demonstrates that , most RPEs remain within $\pm 2\%$, except for the components $L_{20}$. The median values of each box plot are nearly 0, all being less than $\pm 1\%$. Extending the data length $N$ is expected to reduce the median values to 0. Moreover, by adding more excitation signals, the variance of the components $L_{20}$ estimates can be reduced. Table~\ref{tab:7nodesimmersed_defect} indicates that the mean estimates, even for faulty components, are closely approximate to the actual values. This experiment demonstrates that the faulty component $L_{12}$, despite being connected to other faulty components $R_{10}$ and $R_{20}$, can still be detected and identified simultaneously in a subnetwork with only one excitation and partial measurements.


\section{CONCLUSIONS}\label{sec:conclusion and future}
We have introduced a frequency-domain identification method for estimating continuous-time models of diffusively coupled networks. A three-step frequency-domain identification method has been developed that preserves the physical network structure and is used to estimate the values of all physical components in the network. The primary advantages of this method include the direct identification of continuous-time networks and the accurate recovery of the values of physical components in the networks. Furthermore, this method has been extended to identify the local physical components in a subnetwork with partial measurements. A successful application to fault detection and diagnosis in In-Circuit Testing in PCBAs has been shown.


\newpage
\appendix
\section{Appendix}
\subsection{The structure of $Q$ and $\theta$ for full network identification}\label{App: The structure of
Q1 theta1}
\subsubsection{Parameters of the structured network model $\theta$}
The polynomial matrix $A(p)$ has been defined in Definition~\ref{def:diffusively network}, for its $(i,j)$ element defined as $a_{ij}(p) = \sum^{n_a}_{\ell=0} a_{ij,\ell} p^\ell$. The polynomial matrix $B(p)$ and the vector $C(p)$ are defined as $b_{ij}(p) = \sum^{n_b}_{\ell=0} b_{ij,\ell} p^\ell$ and $c_{j}(p) = \sum^{n_c}_{\ell=0} c_{j,\ell} p^\ell$. $\theta$ collects all the parameters into a column vector as $\theta = \left[\begin{array}{ccc}
   \theta_a^T  & \theta_b^T &\theta_c^T
\end{array}\right]^T$, and $\theta_a$, $\theta_b$ and $\theta_c$ are given as follows:

\begin{align}
    &\theta_a = \begin{bmatrix}
             \theta_{a1} \\ \theta_{a2} \\ \vdots \\ \theta_{aL}
    \end{bmatrix}, \quad
    \theta_{ai} = \begin{bmatrix}
              \theta_{a1j} \\ \theta_{a2j} \\ \vdots \\ \theta_{aLj}
    \end{bmatrix}, \quad
    \theta_{aij} = \begin{bmatrix}
             a_{ij,0} \\ a_{ij,1} \\ \vdots \\ a_{ij,n_a}
    \end{bmatrix}, \nonumber \\
    &\text{with } i = 1,..,L \text{ and } j = i,...,L,
    \label{eq: theta_a}
\end{align}

\begin{align}
    &\theta_b = \begin{bmatrix}
             \theta_{b1} \\ \theta_{b2} \\ \vdots \\ \theta_{bL}
    \end{bmatrix}, \quad
    \theta_{bi} = \begin{bmatrix}
              \theta_{b1j} \\ \theta_{b2j} \\ \vdots \\ \theta_{bLj}
    \end{bmatrix}, \quad
    \theta_{bij} = \begin{bmatrix}
         b_{ij,0} \\ b_{ij,1} \\ \vdots \\ b_{ij,n_b}
    \end{bmatrix}, \nonumber \\
    &\text{with } i = 1,..,L \text{ and } j = 1,..,K,
    \label{eq: theta_b}
\end{align}

\begin{align}
    \theta_c &=
\begin{bmatrix}
    \theta_{c1} \\
    \theta_{c2} \\
    \vdots \\
    \theta_{cL}
\end{bmatrix}, \quad
\theta_{cj} =
\begin{bmatrix}
    c_{j,0} \\
    c_{j,1} \\
    \vdots \\
    c_{j,n_c}
\end{bmatrix}, \text{ with } j = 1,..,L,
    \label{eq: theta_c}
\end{align}
where $\theta_a$ is a $\frac{1}{2}L(L+1)(n_a+1)\times1$ vector since $A(p,\theta)$ is in symmetric structure, $\theta_b$ is a $L K(n_b+1)\times1$ vector, and $\theta_c$ is a $L(n_c+1)\times1$ vector.

\subsubsection{The structure of the regressor $Q$}
Define a $L\times L$ complex matrix ${\mathfrak{{W}}}(k)$ for each frequency sample $k$ to collect the frequency weighting part in (\ref{eq: i/o data criterion 2}) as
\begin{equation}
    {\mathfrak{{W}}}(k) = {{\left[ {\hat{C}_W(k)}^{\frac{1}{2}}A{{({{\Omega }_{k}},\theta )}^{i-1}} \right]}^{-1}},
\end{equation}
and $\mathfrak{{W}}_{ij}$ is the $(i,j)$-th element of ${\mathfrak{{W}}}(k)$. Since the polynomial matrix $A(\Omega_k,\theta_a)^{i-1}$ from the last iteration is known, we can substitute the frequency variables $\Omega_k$ into it such that ${\mathfrak{{W}}}(k)$ is obtained by taking the inverse of $A(\Omega_k,\theta_a)^{i-1}$ and pre-multiplying $L \times L$ complex matrix $\hat{C}_W^{-\frac{1}{2}}(k)$.

Define $\Omega_{a}$, $\Omega_{b}$, and $\Omega_{c}$ as vectors that collects the frequency variables of the matrices $A{({{\Omega }_{k}},\theta )}$, $B{({{\Omega }_{k}},\theta )}$, and $C{({{\Omega }_{k}},\theta )}$, respectively,

\begin{equation}
\begin{aligned}
    &\Omega_{a} = \left[\begin{array}{ccccc}
        1 & \Omega_k & \Omega_k^2 &\cdots & \Omega_k^{n_a}
    \end{array}\right], \\
    &\Omega_{b} = \left[\begin{array}{ccccc}
        1 & \Omega_k & \Omega_k^2 &\cdots & \Omega_k^{n_b}
    \end{array}\right], \\
    &\Omega_{c} = \left[\begin{array}{ccccc}
        1 & \Omega_k & \Omega_k^2 &\cdots & \Omega_k^{n_c}
    \end{array}\right].
    \label{eq: omega_abc}
\end{aligned}
\end{equation}

Extracting all the parameters defined as the column vector $\theta$ in (\ref{eq: theta_a}), (\ref{eq: theta_b}) and (\ref{eq: theta_c}) from $M_1(k)$ in \eqref{eq:MQ_theta}, the regressor $Q(k)$ in each sample $k$ is given by,
\begin{equation}
    Q(k) = \left[\begin{array}{ccc}
      \Pi_{A}  &  -\Pi_{B} & -\Pi_{C}
    \end{array}\right].
\end{equation}

$\Pi_{A}$ is a ${L \times \frac{L(L+1)(n_a+1)}{2} }$ matrix given by
\begin{align}
    &\Pi_{A} = \left[\begin{array}{cccc}
     \Phi_{W_{11,1}}\Omega_a & \Phi_{W_{12,1}}\Omega_a &\cdots & \Phi_{W_{LL,1}}\Omega_a \\
     \vdots & \vdots &\vdots &\vdots \\
    \Phi_{W_{11,L}}\Omega_a & \Phi_{W_{12,L}}\Omega_a &\cdots & \Phi_{W_{LL,L}}\Omega_a
    \end{array}\right], \nonumber \\
    & \Phi_{W_{ij,n}} =
    \begin{cases}
    \phi_{{M}_{ii}}, & \text{if } i = j, \\
    (\phi_{M_{ij}} + \phi_{M_{ji}}), & \text{if } i < j,
    \end{cases} \nonumber \\
    &\text{with } i = 1,..,L, \text{ and } j = i,...,L.
\end{align}
$\phi_{M_{ij}}$ denotes the $(i,j)$-th element of $\phi_{M}$, $\phi_M \in \mathbb{C}^{L\times L}$ is obtained by the weightings multiplied by the output samples $W(k)$ vector as $\phi_{M} = \left[
     \mathfrak{W}_{n1} \text{ } \mathfrak{W}_{n2} \text{ } \cdots\text{ }  \mathfrak{W}_{nL}
\right]^T \times \left[
    W_1\text{ }W_2 \text{ }\cdots\text{ } W_L
\right]$ with $n = 1,...,L$.

$\Pi_{B}$ is given as a ${L\times LK (n_b+1)}$ matrix,
\begin{align}
    &\Pi_{B} = \left[\begin{array}{cccc}
     \Phi_{R_{11}} & \Phi_{R_{12}} & \cdots & \Phi_{R_{1L}} \\
     \vdots & \vdots &\vdots &\vdots \\
     \Phi_{R_{L1}} & \Phi_{R_{L2}} & \cdots & \Phi_{R_{LL}}
    \end{array}\right], \nonumber \\
    &\Phi_{Rij} = \left[\begin{array}{cccc}
     \mathfrak{W}_{ij}\Omega_bR_1  & \mathfrak{W}_{ij}\Omega_bR_2 & \cdots & \mathfrak{W}_{ij}\Omega_bR_K
    \end{array}\right], \nonumber \\
     &\text{with } i = 1,..,L, \text{ } j = 1,...,L;
\end{align}
where $R_1$ to $R_K$ is the excitation signal sample on node 1 to node $K$.

$\Pi_{C}$ is given as a ${L\times(n_c+1)}$ matrix,
\begin{equation}
\begin{aligned}
    &\Pi_{C} = \left[\begin{array}{cccc}
     \mathfrak{W}_{11}\Omega_c  & \mathfrak{W}_{12}\Omega_c & \cdots & \mathfrak{W}_{1L}\Omega_c \\
     \vdots & \vdots & \vdots & \vdots \\
      \mathfrak{W}_{L1}\Omega_c  & \mathfrak{W}_{L2}\Omega_c & \cdots & \mathfrak{W}_{LL}\Omega_c
    \end{array}\right]. \\
\end{aligned}
\end{equation}
By using the Kronecker product, $\Pi_{A}, \Pi_{B}$ and $\Pi_{C}$ can be derived as
\begin{equation}
\begin{aligned}
    \Pi_{A} &= \Phi_{W} \otimes \Omega_a, \\
    \Pi_{B} &= \mathfrak{W} \otimes \left[\begin{array}{cccc}
    \Omega_bR_1 & \Omega_bR_2 & \cdots &\Omega_bR_K
    \end{array}\right], \\
    \Pi_{C} &= \mathfrak{W} \otimes \Omega_c.
\end{aligned}
\end{equation}
The dimensions of $Q(k)$ for each frequency sample $k$ is $L\times[\frac{L(L-1)(n_a+1)}{2} +LK (n_b+1)+L(n_c+1)]$. For
the number of $F$ frequency data, $Q$ is collected in $L\times[\frac{L(L-1)(n_a+1)}{2}+LK (n_b+1)+L(n_c+1)]\times F$.

\subsection{Lagrangian multiplier optimization constraints}\label{lagrangian constraint}
We set the constraint for the partially known dynamics or known parameters of the model.
Here, we assume that all the dynamics of the input signals are known, which means that the polynomial matrix $B$ are known. Therefore, the constraint for $B$ as is given by
\begin{equation}
    \Gamma_b = \left[\begin{array}{ccc}
      \mathrm{I}_{n_b+1} \\
      \quad  &  \ddots & \quad \\
      \quad & \quad & \mathrm{I}_{n_b+1}
    \end{array}\right],
\end{equation}
with the dimension $LK(n_b+1)\times LK(n_b+1)$ and $\upsilon_b$ collects all the known parameters of $B$ in the same structure of $\theta_b$ in (\ref{eq: theta_b}). Moreover, if some component dynamics is already known, we can also set the constraint with the known real parameters in the $A$ matrix in the same way. Finally, the constraints $\Gamma$ and $\upsilon$ collect all the above constraints, which are given by,

\begin{equation}
    \Gamma = \left[\begin{array}{ccc}
    \mathrm{O} & \Gamma_b & \mathrm{O}
    \end{array}\right],
    \upsilon = \left[\begin{array}{c}
         \upsilon_b
    \end{array}\right],
\end{equation}
where, $\mathrm{O}$ are the zero matrices with the proper dimension.

\subsection{The structure of $Q_1$ and $\theta_1$ for subnetwork identfication}\label{App: The structure of
Q3 theta3}
\subsubsection{The parameters vector $\theta_1$}
$\theta_1 = \left[\begin{matrix}
    \theta_{Ajj}^T & \theta_{Ajd}^T & \theta_{Bj}^T
\end{matrix}\right]$ The vectors $\theta_{Ajj}$, $\theta_{Ajd}$ and $\theta_{Bj}$ are given by,
\begin{align}
    &\theta_{Ajj} = \begin{bmatrix}
      \theta_{a1} \\ \theta_{a2} \\ \vdots \\ \theta_{aL_{J}}
    \end{bmatrix}, \quad
    \theta_{ai} = \begin{bmatrix}
           \theta_{a1j} \\ \theta_{a2j} \\ \vdots \\ \theta_{aL_{J}j}
    \end{bmatrix}, \quad
    \theta_{aij} = \begin{bmatrix}
            a_{ij,0} \\ a_{ij,1} \\ \vdots \\ a_{ij,n_a}
    \end{bmatrix}, \nonumber \\
    &\text{with } i,j = 1,..,L_{J},
    \label{eq: theta_ajj}
\end{align}

\begin{align}
    &\theta_{Ajd} = \begin{bmatrix}
         \theta_{a1} \\ \theta_{a2} \\ \vdots \\ \theta_{aL_{J}}
    \end{bmatrix}, \quad
    \theta_{ai} = \begin{bmatrix}
        \theta_{a1j} \\ \theta_{a2j} \\ \vdots \\ \theta_{aL_{J}j}
    \end{bmatrix}, \quad
    \theta_{aij} = \begin{bmatrix}
        a_{ij,0} \\ a_{ij,1} \\ \vdots \\ a_{ij,n_a}
    \end{bmatrix}
      , \nonumber \\
    &\text{with } i = 1,..,L_{J} \text{ and } j = L_{J}+1,..,L_{S},
    \label{eq: theta_ajd}
\end{align}

\begin{align}
    &\theta_{Bj} = \begin{bmatrix}
         \theta_{b1} \\ \theta_{b2} \\ \vdots \\ \theta_{bL_{J}}
    \end{bmatrix}, \quad
    \theta_{bi} = \begin{bmatrix}
              \theta_{b1j} \\ \theta_{b2j} \\ \vdots \\ \theta_{bL_{J}j}
    \end{bmatrix}, \quad
    \theta_{bij} = \begin{bmatrix}
              b_{ij,0} \\ b_{ij,1} \\ \vdots \\ b_{ij,n_b}
    \end{bmatrix}, \nonumber \\
    &\text{with } i = 1,..,L_{J} \text{ and } j = 1,..,K,
    \label{eq: theta_bj}
\end{align}
where $L_J\times L_J$ is the dimension of the target subnetwork matrix $A_{JJ}$, $L_J\times L_S$ is dimension of $[A_{JJ}\text{ } A_{JD}]$, $\theta_{Ajj}$ is a vector $L_{J}^2(n_a+1)\times1$, $\theta_{Ajd}$ is a vector $L _{J}L_{D}(n_a+1)\times1$ with $L_D = L_S - L_J$ and $\theta_{Bj}$ is a vector $L_{J}K(n_b+1)\times1$. Collecting the parameters $\theta_{Ajj}$ and $\theta_{Ajd}$ in the matrix $A$ independently can make the construction of the regressor matrix more convenient.

\subsubsection{The structure of the regressor $Q_1$}
The definition of the element $(i,j)$ for the matrix $A(\Omega_k,\theta_1)$, $B(\Omega_k,\theta_1)$ but separated for $A_{JJ}$ and $A_{JD}$ as,
element defined as,
\begin{align}
       & \Tilde{a}_{J_{ij}}(\Omega_k) = \sum^{n_{a}}_{\ell=0} a_{{JJ}_{ij},\ell} \Omega_k^\ell, \quad
       \Tilde{a}_{D_{ij}}(\Omega_k) = \sum^{n_{a}}_{\ell=0} a_{{JD}_{ij},\ell} \Omega_k^\ell, \nonumber \\
       & \Tilde{b}_{{ij}}(\Omega_k) = \sum^{n_{b}}_{\ell=0} b_{{J}_{ij},\ell} \Omega_k^\ell.
\label{eq: tilde_abc_2}
\end{align}

Define the orders $n_{a_{im}}$ and $n_{b_{im}}$ for the immsered DCN model $A_{im}(p)$ and $B_{im}(p)$, respectively. The estimated (scaled) immersed network $\hat{A}_{im}(\hat{\eta})$ and $\hat{B}_{im}(\hat{\eta})$ in frequency can be defined as $\hat{A}_{im}(\Omega_k,\hat{\eta})$ and $\hat{B}_{im}(\Omega_k,\hat{\eta})$, with theirs $(i,j)$ element defined as,
\begin{equation}
\begin{aligned}
       & \hat{a}_{IMJ_{ij}}(\Omega_k) = \sum^{n_{a_{im}}}_{\ell=0} a_{IMJJ_{ij},\ell} \Omega_k^\ell, \\
      & \hat{a}_{IMD_{ij}}(\Omega_k) = \sum^{n_{a_{im}}}_{\ell=0} a_{IMJD_{ij},\ell} \Omega_k^\ell, \\
       & \hat{b}_{IMJ_{ij}}(\Omega_k) = \sum^{n_{b_{im}}}_{\ell=0} b_{IMJ_{ij},\ell} \Omega_k^\ell,
\label{eq: hat_ab_1}
\end{aligned}
\end{equation}
where $\hat{a}_{IMJ_{ij}}(\Omega_k)$ and $\hat{a}_{IMD_{ij}}(\Omega_k)$ are the estimation frequency response data with each frequency point for each element in the estimated $\hat{A}_{JJ}(\hat{\eta})$ and $\hat{A}_{JD}(\hat{\eta})$, respectively. The first equation in (\ref{eq:immerback2}) can be expressed as.
\begin{equation}
\begin{aligned}
    M_3 &= \left[\begin{matrix}
    \mathfrak{J}_{a11} & \mathfrak{J}_{a12} &\cdots & \mathfrak{J}_{a1L_{D}} \\
    \vdots & \vdots & \vdots & \vdots \\
    \mathfrak{J}_{aL_{J}1} & \mathfrak{J}_{aL_{J}2}& \cdots & \mathfrak{J}_{aL_{J}L_{D}}
    \end{matrix}\right] \\
    &- \left[\begin{matrix}
    \mathfrak{D}_{a11} & \mathfrak{D}_{a11} & \cdots & \mathfrak{D}_{a1L_{D}} \\
    \vdots & \vdots & \vdots & \vdots \\
    \mathfrak{D}_{aL_{J}1} & \mathfrak{D}_{aL_{J}2} & \cdots & \mathfrak{D}_{aL_{J}L_{D}}
    \end{matrix}\right],
\end{aligned}
\end{equation}
where $\mathfrak{J}_{aij}$ and $\mathfrak{D}_{aij}$ are given by
\begin{align}
    &\mathfrak{J}_{aij} = \hat{a}_{IMD_{1j}}\Tilde{a}_{J_{i1}}+\hat{a}_{IMD_{2j}}\Tilde{a}_{J_{i2}}+\cdots+\hat{a}_{IMD_{L_{J}j}}\Tilde{a}_{J_{iL_{J}}}, \nonumber \\
    &\mathfrak{D}_{aij} = \hat{a}_{IMJ_{i1}}\Tilde{a}_{D_{1j}}+\hat{a}_{IMJ_{i2}}\Tilde{a}_{D_{2j}}+\cdots+\hat{a}_{IMJ_{iL_{J}}}\Tilde{a}_{D_{L_{J}j}}, \nonumber \\
    &\text{with }i = 1,...,L_J \text{ and } j = 1,...,L_D.
\end{align}

The second equation in (\ref{eq:immerback2}) can be expressed as
\begin{equation}
\begin{aligned}
    M_4 &= \left[\begin{matrix}
    \mathfrak{J}_{b11} & \mathfrak{J}_{b12} &\cdots & \mathfrak{J}_{b1K} \\
    \vdots & \vdots & \vdots & \vdots \\
    \mathfrak{J}_{bL_{J}1} & \mathfrak{J}_{bL_{J}2}& \cdots & \mathfrak{J}_{bL_{J}K}
    \end{matrix}\right] \\
    &- \left[\begin{matrix}
    \mathfrak{B}_{b11} & \mathfrak{B}_{b11} & \cdots & \mathfrak{B}_{b1K} \\
    \vdots & \vdots & \vdots & \vdots \\
    \mathfrak{B}_{bL_{J}1} & \mathfrak{B}_{bL_{J}2} & \cdots & \mathfrak{B}_{bL_{J}K}
    \end{matrix}\right],
\end{aligned}
\end{equation}
where, $\mathfrak{J}_{bij}$ and $\mathfrak{B}_{bij}$ are given by
\begin{align}
    &\mathfrak{J}_{bij} = \hat{b}_{IMJ_{1j}}\Tilde{a}_{J_{i1}}+\hat{b}_{IMJ_{2j}}\Tilde{a}_{J_{i2}}+\cdots+\hat{b}_{IMJ_{L_{J}j}}\Tilde{a}_{J_{iL_{J}}}, \nonumber \\
    &\mathfrak{B}_{bij} = \hat{a}_{IMJ_{i1}}\Tilde{b}_{J_{1j}}+\hat{a}_{IMJ_{i2}}\Tilde{b}_{J_{2j}}+\cdots+\hat{a}_{IMJ_{iL_{J}}}\Tilde{b}_{J_{L_{J}j}}, \nonumber \\
    &\text{with }i = 1,...,L_J \text{ and } j = 1,...,K.
\end{align}

Extracting all the parameters defined as the column vector $\theta_1$ in (\ref{eq: theta_ajd}), (\ref{eq: theta_ajd}) and (\ref{eq: theta_bj}) from $M_3$ and $M_4$, the regressor $Q_1$ is given by
\begin{equation}
    Q_1 = \left[\begin{array}{cc}
      \Pi_{A1}  &  \Pi_{B1}
    \end{array}\right]^T,
\end{equation}
with
\begin{align}
    \Pi_{A1} &= \left[\begin{array}{cc}
      \Pi_{A11}  &  -\Pi_{A12}
    \end{array}\right], \nonumber \\
    \Pi_{A11} &= \mathrm{I}_{L_J} \otimes (\hat{A}_{JD}(\Omega_k,\hat{\eta})^T \otimes \Omega_a),  \\
    \Pi_{A12} &= \hat{A}_{JJ}(\Omega_k,\hat{\eta}) \otimes (\mathrm{I}_{L_D} \otimes \Omega_a), \nonumber
\end{align}
and
\begin{equation}
\begin{aligned}
    \Pi_{B1} &= \left[\begin{array}{cc}
      \Pi_{B11}  &  -\Pi_{B12}
    \end{array}\right], \\
    \Pi_{B11} &= \mathrm{I}_{L_J} \otimes (\hat{B}_{J}(\Omega_k,\hat{\eta})^T \otimes \Omega_a), \\
    \Pi_{B12} &= \hat{A}_{JJ}(\Omega_k,\hat{\eta}) \otimes (\mathrm{I}_{K} \otimes \Omega_b). \\
\end{aligned}
\end{equation}

The dimensions of $Q_1(k)$ for each frequency sample $k$ is $[L_J(L_D+K)]\times [L_JL_S(n_a+1)+L_JK(n_b+1)]$, and for
the number of $F$ frequency data, $Q_1$ can be collected in $[L_J(L_D+K)]\times [L_JL_S(n_a+1)+L_JK(n_b+1)]\times F$.

\bibliographystyle{ieeetr}
\bibliography{references}


\end{document}